\documentclass[prd, showpacs,preprintnumbers,amsmath,amssymb, nofootinbib]{revtex4}
\usepackage{amsmath,amssymb,graphicx}
\usepackage{epsf}
\usepackage{epstopdf}
\usepackage {amssymb}
\usepackage{color}
\newcommand{\nc}{\newcommand}
\nc{\ba}{\begin{eqnarray}}
\nc{\ea}{\end{eqnarray}}
\nc{\calR}{{\cal R}}
\nc{\calO}{{\cal O}}
\newcommand\be{\begin{equation}}
\newcommand\ee{\end{equation}}

\newcommand\bx{{\mathbf{x}}}
\newcommand\bk{{\mathbf{k}}}

\begin{document}

\title{Asymmetric Sky from the  Long Mode Modulations}

\author{Ali Akbar Abolhasani$^{1}$}
\email{abolhasani-AT-ipm.ir}
\author{Shant Baghram$^{2}$}
\email{baghram-AT-ipm.ir}
\author{Hassan Firouzjahi$^{2}$}
\email{firouz-AT-mail.ipm.ir}
\author{Mohammad Hossein Namjoo$^{1}$}
\email{mh.namjoo-AT-ipm.ir}

\affiliation{$^1$School of Physics, Institute for Research in
Fundamental Sciences (IPM),
P.~O.~Box 19395-5531,
Tehran, Iran}

\affiliation{$^2$School of Astronomy, Institute for Research in
Fundamental Sciences (IPM),
P.~O.~Box 19395-5531,
Tehran, Iran}

\begin{abstract}
\vspace{0.3cm}
The observed dipole asymmetry in Cosmic Microwave Background radiation
may have originated from the modulations of super-horizon long wavelength modes.
In this work we unveil different aspects of asymmetries generated from the long wavelength mode modulations. We show that the same  mechanism  which leads to  the observed CMB power spectrum dipole asymmetry from the long mode modulations also yields dipole asymmetry in (a): tensor perturbations power spectrum and  (b): the halo bias parameter.
These are different phenomena relevant to different cosmological histories but both share the same underlying mechanism in generating asymmetries in the sky. We obtain the set of consistency conditions relating  the amplitude of dipole asymmetries generated on tensor perturbations and  halo bias parameter to  the amplitude of dipole asymmetry generated on CMB power spectrum. In addition, we show that this mechanism does not produce dipole asymmetry in acceleration expansion in $\Lambda CDM$ universe because the super-horizon curvature perturbation is conserved in this background.

\vspace{0.3cm}

\end{abstract}

\date\today

\maketitle

\section{Introduction}
In theories of  early Universe it is conceivable that our observable Universe is a patch of much larger Universe. For example, the inflationary period \cite{Guth:1980zm} does  not  necessarily last only $60$ e-folds or so to solve minimally the horizon and the flatness problems today. But it is natural to imagine that inflation last longer. Indeed, we can think about pre-inflationary effects on our observable Universe. Two main pre-inflationary effects are bubble collisions and long mode modulations. Historically,  Grishchuk and Zel'dovich were the first who proposed that the  super-horizon perturbations can generate large-scale temperature fluctuations on the cosmic microwave background (CMB) \cite{Grishchuk:1987}.

The idea of looking for  the fingerprints  of the pre-inflationary physics in sky
is boosted again with the  detection of hemispherical asymmetry on CMB as reported by the   Planck collaboration  \cite{Ade:2013nlj} which was also observed  by Wilkinson Microwave Anisotropy Probe (WMAP) data \cite{Eriksen:2003db}. In the case that the observed anomaly is not a statistical artifact, it can be viewed  as a new challenge for simple single field inflationary models \cite{Dai:2013kfa,{Pullen:2007tu}}. In this venue Erickcek et al \cite{Erickcek:2008sm} and  Gordon\cite{Gordon:2006ag} argued that the modulation of the super-horizon perturbations can be the source of this asymmetry. However this modulations must be treated carefully, since it can produce large temperature perturbations on CMB map on which there are  strict constrains  on the departure from isotropy (mainly from quadrupole and octupole moments ) \cite{Erickcek:2008jp}.


The aim of this  letter is not to study the fundamental origin of this long mode modulation.  Instead, assuming the existence of this large amplitude long wavelength mode, we would like to examine the consequences of such long mode modulations on various cosmological parameters. In particular we examine whether or not  the same mechanism which generates dipole asymmetry on CMB power spectrum can also generate dipole asymmetry on (a): tensor perturbations, (b): halo bias parameter and (c): dark energy acceleration expansion. Our findings show that the long modulation induce dipole asymmetry, at least in principle,
in cases (a) and (b) but  not in case (c) in a $\Lambda CDM$ background. We also relate the
amplitude of dipole asymmetry generated in tensor perturbations and  halo bias parameters  to  the amplitude of dipole asymmetry generated on CMB power spectrum.  In this view, our study provides a set of consistency conditions for the asymmetries generated on seemingly different cosmological observables and histories which can be tested in future observations, see also \cite{Dai:2013kfa} for a similar line of thought.
Having this said, recently the possibility of exciting long mode perturbations in open inflation models  via bubble nucleation was put forward in \cite{Liddle:2013rta} which is very intriguing.

A phenomenological parameterizations of the bipolar asymmetry is defined via
\ba
\label{P-asym}
{\cal P}^{1/2}_{\cal R}(k, {\bf x}) = {\cal P}^{1/2^{iso}}_{{\cal R}}(k) (1+ A(k) {\bf \hat p.x}/x_{CMB} )
\ea
in which $\calR$ is the comoving curvature perturbation,
${\cal P}_{\cal R}(k, {\bf x})$ is the asymmetric curvature perturbations power spectrum, ${\cal P}^{{iso}}_{\cal R}$ is the   isotropic power spectrum, $A(k)$ is the amplitude of the bipolar asymmetry, and the direction of anisotropy is shown by $\bf \hat p$ and   $x_{CMB}$ is the comoving distance to the surface of last scattering. The recent  data from Planck mission indicates that the amplitude of the asymmetry is $A =0.072 \pm 0.022$  for large angular scales, $\ell < 64$.  Data analysis shows that  the best fit for the anisotropy direction is $(l,b)=(227,-27)$ \cite{Ade:2013nlj}.

As mentioned above, beside the observational effects on the CMB temperature fluctuations, the super-horizon long mode will also introduce non-trivial effects on the early and late time cosmological observables.
In  Sec. (\ref{SecAsymm}) we formulate the  asymmetry induced from a long mode perturbations for a general field fluctuations  and apply this formalism for asymmetry generated in tensor perturbations.  In Sec. (\ref{bias}) we focus on the modulation of halo bias parameter and in Section (\ref{decel}) we study the effect of long mode on deceleration parameter
followed by discussions and conclusions in Section (\ref{conclusion}). We present the
technical analysis of calculating the cross-correlation between one scalar and two graviton perturbations in non-attractor backgrounds in  the Appendix.


\section{Asymmetry on a general field fluctuation by large scale perturbation}\label{SecAsymm}

In this section we investigate how a long wavelength scalar perturbation can modulate the
 statistics of a general field $\calO$ on smaller  CMB-scales, yielding a hemispherical asymmetry on its power spectrum ${\cal P}_{\calO}$. The field of interest, $\calO$, is kept arbitrary but we
are mainly interested in the cases where $\calO$ represents the curvature perturbations $\calR$
or the tensor perturbations. In particular, when ${\cal O}=\calR$, it is shown in \cite{Namjoo:2013fka} that the  amplitude of the dipole asymmetry in curvature perturbation power spectrum, $A_\calR$, is proportional to the amplitude of local non-Gaussianity $f_{NL}^{loc}$.  In deriving this conclusion it was essential that only one field sources the curvature perturbation.

With this discussion in mind, now we extend the analysis in \cite{Namjoo:2013fka} for general
${\cal O}$. Similar to the single source assumption employed in \cite{Namjoo:2013fka},
in what follows we assume that only one field, say $\delta \phi$,   has a non-negligible three point cross correlation with  $\calO$, $\langle \delta \phi {\cal O} {\cal O} \rangle \neq 0$.  That is, any other large scale field, $\delta \sigma_i, i=1,2...$, has negligible correlation in the form of $\langle {\delta \sigma_i O O} \rangle$, yielding negligible  modulations \footnote{Actually we need the  weaker requirement  that $\langle { \delta \sigma_i O O} \rangle \rightarrow 0 $
in the squeezed limit $k_L \rightarrow 0$.}.
Under the above assumptions, and as long as the above correlation is concerned, the long wavelength mode is effectively the comoving curvature perturbation $\calR_L$. In other words, the information in three point function  $\langle \delta \phi {\cal O} {\cal O} \rangle$ is encoded in $\langle \calR {\cal O} {\cal O} \rangle$.  This can be seen if one writes down $\calR$ in terms of $\delta \phi$ and $\delta \sigma_i$ perturbations:   $\calR = c_\phi \delta \phi + \sum_i c_i \delta \sigma_i$ with some coefficients $c_\phi$ and $c_i$.

We are interested in asymmetry generated on power spectrum of the operator $\calO$
defined in Fourier space via
\ba
\label{R-power}
\langle \calO_{\bf {k}} \calO_{\bf{k'}} \rangle  = (2 \pi)^3\delta^3  ( {\bf{k} + \bf{k'}})  P_{\calO}(k)
\quad , \quad
{\cal P}_{\calO} \equiv \frac{k^3}{2 \pi^2}  P_{\calO}(k) \, .
\ea
Following the parameterizations given in Eq. (\ref{P-asym}), the dipole asymmetry in the power spectrum of $\calO$ in Fourier space can be modeled by
\ba
{\cal P}_\calO(k) {\simeq}   {\cal P}_\calO^{iso}(k) \left( 1+ 2 A_\calO(k) \hat {\bf p} . {\bf x}/x_n  \right) \, ,
\ea
where ${\cal P}_\calO^{iso}(k)$ is the isotropic part of the power spectrum  and $A_\calO(k)$ represents the amplitude of the dipole asymmetry which we are interested in.  By the above parameterizations, we can write
\ba
\label{grad}
\frac{\nabla P_{{\calO_k}}}{P_{{\calO_k}}} \simeq \frac{2 A_\calO \hat {\bf p}}{x_n}
\ea
which will be used later.

We assume that there exists a large super-horizon mode $\calR_L$ with the amplitude ${\cal P_R}_L$ and the comoving wave number $k_L$, super-imposed on the entire observable Universe
\ba
\label{RL}
\calR_{L} = \calR_{k_L} \sin({\bf k}_L.{\bf x})  = {\cal P_{\calR}}_L^{1/2} \sin({\bf k}_L.{\bf x}) \, .
\ea
Note that ${\cal P_{R}}_L$ is the power spectrum of the long mode obtained via ensemble averaging  in a very large box \cite{Lyth:2007jh}.  In this view, for small scale perturbations inside this very large box, $k \gg k_L$, the quantum fluctuations of $\calO$ are treated as random statistical variables. In this view the size of our observed Universe is given by $H_0^{-1}$ in which $H_0$ is the current Hubble constant but the long mode which causes the modulation has the wavelength $\lambda_L \gg H_0^{-1}$. As usual, for our small CMB-scale modes $k$,  we work in the Fourier space with the volume $V$.  For this picture to work, the volume of the Fourier space  should be bigger than $H_0^{-1}$ but smaller than $\lambda_L$ so we have the following hierarchy in mind
\ba
H_0^{-3} < V \ll k_L^{-3}.
\ea

Now let us parameterize the three point cross correlation function
$\langle \calR({\bf k}_L) \calO( {\bf k_1}) \calO( {\bf k_2}) \rangle$
in the squeezed limit  $k_L \to 0$ by
\ba
\label{fNL}
\langle \calR({\bf k}_L) \calO( {\bf k_1}) \calO( {\bf k_2}) \rangle
\equiv (2 \pi)^3 \delta({\bf{k_L+k_1+k_2}})
\left( \frac{12}{5} f_{NL}^{\calR \calO}  \right)\, P_{\calR}(k_L) P_\calO(k_1)
\ea
In this view, the parameter  $f_{NL}^{\calR \calO}$ measures the three point cross  correlations of $\calR$ and $\calO$, which is a generalization of the usual local non-Gaussianity parameter $f_{NL}^{loc}$. Note that for ${\cal O}=\calR$, we have  $f_{NL}^{\calR \calR} = f_{NL}^{loc}$.

The essential point to note is that the effect of the long wavelength curvature perturbation on small scale perturbations is just a rescaling of their background. This is because, the small CMB-scale perturbations can not probe the spatial variation associated with the long wavelength perturbations $k_L$. Consequently, we expect that the cosmological observations, which only probe the cosmic background will not be affected by long mode modulation.  Keeping in mind  that the curvature perturbation is not necessarily constant on super-horizon scales, this yields \cite{Namjoo:2013fka}
\ba
 \label{rescale}
\Big \langle \calR({\bf k_L}) \calO( {\bf k_1}) \calO( {\bf k_2}) \Big \rangle
&\simeq &
\Big \langle \calR({\bf k_L}) \langle \calO( {\bf k_1}) \calO( {\bf k_2}) \rangle_{\calR({\bf k_L})} \, \Big \rangle
\\ \nonumber
&\simeq &
\Big \langle \calR({\bf k_L}) \,\, \left(   \calR({\bf k_L}) \dfrac{\partial}{\partial \calR({\bf k_L})} \Big \langle \calO({\bf k_1})\calO({\bf k_2}  ) \Big \rangle \large
+  \dot\calR({\bf k_L}) \dfrac{\partial}{\partial \dot\calR({\bf k_L})} \Big \langle \calO({\bf k_1})\calO({\bf k_2}) \Big \rangle \large   \,  \right)
 \Big\rangle
\ea
Here $ \langle \calO( {\bf k_1}) \calO( {\bf k_2}) \rangle_{\calR({\bf k_L})}$ means that we calculate $ \langle \calO( {\bf k_1}) \calO( {\bf k_2}) \rangle$ in the background of
$\calR({\bf k_L})$.   To obtain the above relation we have assumed the Bunch-Davies initial condition  so the non-Gaussianity effects deep inside the horizon  are negligible. Furthermore, as emphasized at the beginning of this Section, it was essential that only one field had non-zero three-point correlation with ${\cal O}$ so $\langle \calO( {\bf k_1}) \calO( {\bf k_2}) \rangle$ depends on the modulations of a single  field which can be absorbed in $\calR(k_L)$ as we did in Eq. (\ref{rescale}).

Now comparing Eq. \eqref{fNL} with Eq. \eqref{rescale} one has
\ba
\label{squeeze-dotzeta}
\dfrac{12}{5} f_{NL}^{\calR \calO} P_{\calR_L} P_{\calO} \simeq
P_{\calR_L} \dfrac{\partial P_\calO}{\partial \calR_L}  + \dfrac{1}{2} \partial_t{P_{\calR_L}} \dfrac{\partial P_\calO}{\partial \dot \calR_L} \, .
\ea
On the other hand, if  $\calR_L$ is responsible for the asymmetry of the power spectrum of field $\calO$ we have
\ba
\label{eq1}
\nabla P_\calO =
 \dfrac{\partial P_\calO}{\partial \calR_L} \nabla \calR_L + \dfrac{\partial P_\calO}{\partial \dot \calR_L} \nabla \dot{\calR_L}
\ea
Noting that {in coordinate space}, $\nabla \calR_L = k_L \calR_L$ and $\calR_L = {\cal P}_{\calR_L}^{1/2}$ we have
\ba
\label{eq2}
\calR_L \nabla P_\calO =
 \dfrac{\partial P_\calO}{\partial \calR_L} k_L {\cal P}_{\calR_L} + \dfrac{1}{2}\dfrac{\partial P_\calO}{\partial \dot \calR_L}
 k_L  \dot{\cal P}_{\calR_L} \, .
\ea
Comparing this with \eqref{squeeze-dotzeta} yields
\ba
\label{eq3}
\dfrac{\nabla P_\calO}{P_\calO}
\simeq
\dfrac{12}{5} f_{NL}^{\calR \calO}  \, {\bf k}_L  {\cal P}_{\calR_L}^{1/2} .
\ea
Finally, using \eqref{grad} one obtains
\ba
\label{A-O}
A_\calO \simeq \dfrac{6}{5} f_{NL}^{\calR \calO}  \,x_n k_L  {\cal P}_{\calR_L}^{1/2} \, .
\ea
This is one of the main result for this section. This formula relates $A_\calO $,
the amplitude of the dipole asymmetry in $P_\calO$, to the cross term coupling $f_{NL}^{\calR \calO}$ and the amplitude of the long mode perturbations
${\cal P}_{\calR_L}^{1/2}$.

Now we can use Eq. (\ref{A-O})  to obtain the amplitude of modulation for some interesting examples.

\subsection{CMB Power Spectrum Dipole Asymmetry }

The first example corresponds to the case $\calO=\calR$ so we can calculate the dipole asymmetry in CMB power spectrum similar to
\cite{Namjoo:2013fka, Lyth:2013vha, Wang:2013lda, Erickcek:2009at, McDonald:2013aca, Mazumdar:2013yta, Liu:2013kea}, see also \cite{Schmidt:2012ky, Prunet:2004zy, Byrnes:2011ri}.
With $f_{NL}^{\calR \calR} = f_{NL}^{loc}$, the amplitude of the CMB dipole
asymmetry,   $A_{\calR}$,  from Eq. (\ref{A-O}) is obtained to be
\ba
\label{AR}
A_\calR \simeq \dfrac{6}{5} f_{NL}^{loc}  \,x_n k_L  {\cal P}_{\calR_L}^{1/2}  \, .
\ea

In order to obtain observable bipolar asymmetry from the long mode modulation one needs
$ {\cal P}_{\calR_L} \gg  {\cal P}_{\calR}(k_{CMB})$. However,  in order  for perturbations to be under control we require $ \calR_L^2  \simeq   {\cal P}_{\calR_L}(k_L) \lesssim 1$.
On the other hand, imposing the octupole $Q_3$ constraints on CMB anisotropies   \cite{Erickcek:2008jp, Namjoo:2013fka} yields
\ba
\label{bound3}
\frac{6}{5}(k_L x_n) \,  {\cal P}_{\calR_L}^{1/2} \lesssim  {32Q_3}^{1/3} \sim 10^{-1} \, .
\ea
Plugging Eq.  (\ref{bound3}) in Eq. (\ref{AR}) we obtain our upper bound consistency condition  for the amplitude of the CMB dipole  asymmetry \cite{Namjoo:2013fka}
\ba
\label{A-upper}
| A_{\calR} |  \lesssim 10^{-1} | f_{NL}^{loc} | \, .
\ea

However, it is pointed out in \cite{Lyth:2013vha, Erickcek:2008jp} that there is another term in the total curvature perturbation proportional to $\calR_L^2$. This brings  another constraint from the quadrupole $Q_2$ on CMB. This changes the above bound to \cite{Lyth:2013vha, Erickcek:2008jp}
\ba
\label{A-upper-lyth}
 | A_{\calR} |  \lesssim 0.02 \, | f_{NL}^{loc} |^{1/2}
\ea
See, however, \cite{Kanno:2013ohv} for a possible way to avoid this bound.
In order to obtain observable dipole asymmetry consistent with the Planck observation we need
$\vert A_{\calR} \vert =0.07 \pm 0.02$.

Eq. (\ref{A-upper-lyth}) is an interesting result, relating the amplitude of the hemispherical asymmetry to the level of non-Gaussianity in the system. This was first obtained by Lyth in \cite{Lyth:2013vha} for the conventional curvaton scenario in which entire curvature perturbation is sourced by the curvaton field.
For the single field attractor models of inflation with the Bunch-Davies initial condition
in which the level of non-Gaussianity is related
to the spectral index $n_s$ via \cite{Maldacena:2002vr} $f_{NL}^{loc} \sim n_s-1$, Eq. (\ref{A-upper-lyth}) yields  $| A_{\calR} |  \lesssim 10^{-2} (n_s-1)^{1/2}
\sim 10^{-3}$ which is too small to be observable.

As speculated in \cite{Namjoo:2013fka}, one can obtain large hemispherical asymmetry in  models of non-attractor single field inflation in which the curvature perturbation is not frozen on
super-horizon scale and large observable non-Gaussianity can be generated independent of
$n_s$ via \cite{Namjoo:2012aa, Chen:2013aj, Huang:2013oya, Chen:2013kta}
\ba
\label{f-NLcs}
f_{NL}^{loc} =\frac{5(1+ c_s^2)}{4 c_s^2} \, ,
\ea
where $c_s$ is the sound speed of cosmological perturbations during the non-attractor phase.
With $f_{NL}^{loc} \sim \mathrm{few}$ one can saturate both the Planck
constraints  on local non-Gaussianity \cite{Ade:2013ydc} and the observed CMB bipolar asymmetry. As discussed in \cite{Namjoo:2013fka}, Eq. (\ref{f-NLcs}) holds only during the short non-attractor phase which precedes the last 60 or so e-folds of inflation. As a result, once the attractor phase of inflation has reached, one obtains the usual relation $f_{NL} \sim n_s-1$ so the amplitude of bipolar asymmetry on smaller CMB scales rapidly becomes negligible. This built-in scale-dependence of $f_{NL}$ can address the  quasar constraints on $A_{\cal R}$ on scales smaller than $Mpc^{-1}$ \cite{Hirata:2009ar}.

\subsection{Tensor Perturbations  Asymmetry }
\label{tensor}

As a second interesting example, we obtain the amplitude of the dipole asymmetry in  tensor perturbations power spectrum,  $A_T$, induced by the long wavelength mode.

Consider the three-dimensional spatial metric on the surface of constant time
\cite{Maldacena:2002vr}
\ba
ds_{(3)}^2 = a(t)^2 e^{2 \calR(t, \bx)} \,   {^{(3)}}g_{ij} d \bx^i d \bx^j
\ea
in which
\ba
 {^{(3)}}g_{ij}= \delta_{ij} + h_{ij}+ \frac{1}{2} h_{i l} h_{lj} +... \quad,  \quad
  \partial_i h_{ij}= h_{ii}=0 \, .
\ea
With this convention $\det {^{(3)}}g_{ij} =1 $ and $h_{ij}$ represents the two degrees of freedom associated with the tensor perturbations.

Setting $\calO = h_{ij}$ in Eq. (\ref{fNL}) and parameterizing the cross correlation function $\langle \calR h_{ij} h_{kl} \rangle $ by $f_{NL}^{\calR h}$, the amplitude of tensor perturbations dipole asymmetry,  $A_{T}$, from Eq. (\ref{A-O}) is obtained to be
\ba
\label{AT}
A_T \simeq \dfrac{6}{5} f_{NL}^{\calR h}  \,x_n k_L  {\cal P}_{\calR_L}^{1/2} \, .
\ea
This is an interesting result:  \textit{ if there is an enhanced large scale curvature perturbation that modulates the CMB curvature perturbations, it can also modulate the tensor perturbations power spectrum}. The amplitude of this enhancement is controlled by
$f_{NL}^{\calR h}$ which measures the cross-correlation
$\langle \calR h_{ij} h_{kl} \rangle $.  Comparing the amplitude of tensor modulation in
Eq. (\ref{AT}) to the corresponding scalar perturbation modulation $A_{\calR}$ given in
Eq. (\ref{AR}) we obtain the following consistency condition
\ba
\label{AT-AS}
\dfrac{A_T}{A_\calR} \simeq \dfrac{f_{NL}^{\calR h} }{f_{NL}^{loc}} \, .
\ea
Note that the above relation is valid for all single field inflationary models with the Bunch-Davies initial condition, including non-attractor models.
It is worth to mention that the tensor perturbations freeze out after horizon crossing independent of  the model of inflation. As a result, any correlation and modulation for tensor field occurs during inflation and mainly at the time of  horizon crossing. However, depending on the model, the curvature perturbation can evolve even after inflation and still has a chance to have correlation with other fields. This is particularly the case in models of  multiple fields inflation such as in curvaton scenario in which the curvature perturbations are generated at or after the end of inflation by light fields other than the inflaton  field.  Hence, Eq. (\ref{AT-AS}) does not work for e.g. curvaton model. In such models, the curvaton field does not contribute to inflationary phase, neither at the background level nor to perturbations. As a result the tensor perturbation can correlate only with inflaton field at horizon crossing. However, after inflation, this is the curvaton field that has the main role in perturbations and modulating the curvature perturbation.

Let us consider the models of single field slow roll inflation. As mentioned before, we know that these models can not generate the observed CMB dipole asymmetry, i.e. for these models
$A_\calR \ll 0.07$. However we consider these models as  a platform to demonstrate the implication of our consistency condition Eq. (\ref{AT-AS}) as a proof of concept.  From Maldacena's analysis \cite{Maldacena:2002vr},  $f_{NL}^{loc}$
and $f_{NL}^{\calR h}$ are related to the curvature perturbations spectral index $n_s$ and to the tensor perturbation spectral tilt $n_T$ via
\ba
\label{Maldacena}
\dfrac{12}{5} f_{NL}^{loc} = -(1-n_s) \qquad , \qquad   \dfrac{12}{5} f_{NL}^{\calR h} = 2 \epsilon= -n_T \, ,
\ea
in which $\epsilon = -\dot H/H^2 $ is the slow-roll parameter and $H$ is the Hubble expansion rate.   Also note that $n_T$ is related to the ratio of the
amplitude of the tensor perturbations to the amplitude of the scalar perturbations via the consistency condition  $r =  16 \epsilon=  - 8 n_T$. As a result, Eq. (\ref{AT-AS}) yields
\ba
\label{A-O-R}
\big\vert \dfrac{A_T}{A_\calR} \big\vert \simeq  \dfrac{n_T}{n_s-1}
= \frac{r}{8 ( 1- n_s)} \, .
\ea
Observationally this is a very interesting result in which the ratio of the dipole asymmetries in tensor perturbations and the scalar perturbations is given by the ratio of $n_T$ and $1-n_s$.
This is a hybrid of the two consistency relations $f_{NL}^{loc} \sim 1- n_s$ and $r = - 8 n_T$. As an estimation of $A_T$,  using  the Planck constraints $r\lesssim 0.13$ and $n_s \simeq 0.96$, from Eq. (\ref{A-O-R}) we obtain $A_T \lesssim 0.4 A_\calR $. With the  observational bound $A_\calR \simeq 0.07$ this yields the prediction  $A_T \lesssim 0.03$. It is interesting to see whether or not the upcoming analysis of the CMB polarization data by the Planck team   can detect this level of hemispherical asymmetry on tensor perturbations.

As mentioned above, Eq. (\ref{A-O-R}) holds for the single field attractor models with the initial Bunch-Davies vacuum  in which the
Maldacena's consistency relations, Eq. (\ref{Maldacena}), are at work. However, as mentioned above, simple single field models of inflation can not produce large enough dipole asymmetry,
i.e. $A_{\calR} \ll 0.07$ in these scenarios. Therefore, single field slow-roll models
predict $A_T \ll 0.03$. This is too small to be detected observationally.
Therefore, one has to look for alternatives.

As shown in \cite{Namjoo:2013fka} and in previous
subsection,  the single field non-attractor  models which violate Maldacena's consistency condition are able to produce large observable dipole asymmetry as given in Eq. (\ref{f-NLcs}). Therefore, it is an interesting question to calculate the correlation function
$\langle \calR h^2\rangle$ directly in non-attractor models to obtain $f_{NL}^{\calR h}$.
This analysis were performed in Appendix  \ref{App1}. The shape of the bispectrum
as given in Eq. (\ref{shape-full}) is very different than the results obtained by Maldacena.
There are two reasons for this. First, since $\calR$  evolves on super-horizon scales the profile of  its wave function  is different than that of the usual attractor model. This has to be taken into account when calculating the in-in integrals.
Second, in the process of field redefinition (see the Appendix for details), unlike in Maldacena's analysis, one can not neglect terms containing $\dot \calR$.    However, in the squeezed limit in which $k_L \ll k_1 \simeq k_2$, the shape function in
Eq. (\ref{shape-full}) collapses to the result obtained by Maldacena in the attractor phase as given in  Eq. (\ref{shape-squeezed}). As a result, the relation $f_{NL } \sim \epsilon$ as given in Eq. (\ref{Maldacena}) still holds for non-attractor models. However, we note that the relation between $n_s-1$ and $f_{NL}^{loc}$ as given in Eq. (\ref{Maldacena})
does not hold in non-attractor model and one has to use Eq. (\ref{f-NLcs}).  Using the expressions for $f_{NL}^{\calR h}$ and $f_{NL}^{loc}$  respectively from Eqs. (\ref{Maldacena}) and (\ref{f-NLcs}),  we obtain
\ba
\label{AT-AR2}
\frac{A_T}{A_{\calR}} \simeq  \frac{2 \epsilon c_s^2}{3 (1 + c_s^2)}=
\frac{c_s\,  r}{24 (1 + c_s^2) } \, .
\ea
in which in the second equality the relation $r= 16 c_s \epsilon$ has been used.
This equation should be compared with  Eq. (\ref{A-O-R}) obtained for the attractor models.

In models of non-attractor inflation  $\epsilon $ decays like $\epsilon \propto a^{-6}$ \cite{Namjoo:2012aa}  so at the end of non-attractor phase $\epsilon$ and $r$ becomes exponentially small (assuming the non-attractor phase has a few number of e-foldings). This means that the amplitude of tensor perturbations are very small in non-attractor model \cite{Chen:2013aj}. Therefore, the ratio $A_T/A_R$ in non-attractor phase is much smaller than the corresponding value in
conventional slow-roll models. Taking $r \lesssim 0.1$ and $c_s \lesssim 1$,  we obtain $A_T \simeq 3 \times 10^{-4}$.  It is unlikely that the future cosmological observations can detect such small dipole asymmetry in tensor perturbations power spectrum.

To summarize, here we have shown that the long mode modulation induces dipole asymmetry not only on curvature perturbations power spectrum but also on tensor mode perturbations power spectrum.  However, none of the single field models studied so far can generate detectable dipole asymmetry on tensor power spectrum. For single field slow-roll model $A_T \ll 0.03$ because  $A_\calR$ generated in these models
is too small to match the observed value. On the other hand for the non-attractor models $A_T$ is even smaller than this value  because in these models the amplitude of tensor perturbations are suppressed.  As a result, in order to obtain detectable dipole asymmetry in tensor perturbations power spectrum, one has to look for multiple fields models in which more than one field contribute to the curvature perturbations.
In multiple fields  scenarios  the specific formulas Eqs. (\ref{A-O-R}) and (\ref{AT-AR2})
do not hold. Therefore, for a given multiple fields inflation  model, one has to explicitly calculate $\langle \calR h^2\rangle$ and correspondingly obtain the amplitude of $f_{NL}^{ \calR h} $ and see whether large enough $A_T$ is generated.

\section{Modulated Bias}
\label{bias}

In this section we show how the long wavelength modulation ${\cal{R}}_L$ from curvature perturbation can affect the distribution of galaxies in sky through the halo bias parameter. The large scale structure observables deals with the statistics of luminous matter, (i.e galaxies and cluster of galaxies) sitting in the center of dark matter halos \cite{Percival:2006gt}. The statistics of galaxies such as the correlation functions, power spectrum and the probability distribution function are related to the dark matter halo statistics. With the assumption that each dark matter halo is a host of a galaxy \cite{Mo:1995cs} this relation is parameterized through bias parameter $b=\delta_h/\delta_m$ where $\delta_h$  and $\delta_m$ are  the halo density perturbation  and the dark matter density perturbation respectively.   In order to show the effect of long wavelength modulation, we study its effect on the threshold of non linear structure formation. In spherical collapse \cite{Gunn:1972sv}, a region of $R$  undergoes a gravitational instability when its density perturbation becomes greater than the critical density $\delta_c \simeq 1.68$.  This is obtained by considering the evolution of an over-dense region in cosmological background. The over-dense region behaves like a closed Friedmann Universe with spatial curvature  $\Delta k$ and density of $\bar{\rho}_M+\Delta\rho_M$, where $\bar{\rho}_M$ is the background density and $\Delta\rho_M$ is the excess density. The critical density $\delta_c=\Delta\rho_{M(c)}/\bar{\rho}_M$ is related to the critical excess of mass $\Delta\rho_{M(c)}$, where the structure enters the non-linear regime.  

In a recent work \cite{Baghram:2013lxa}, it was shown that the bias parameter is a potential important observable in LSS to detect the primordial anisotropy through the non-linear effect (non-Gaussianity). In this work we just investigate the effects of anisotropy induced from the  long wavelength mode in linear regime due to its effect on the critical density.

\begin{figure}
\includegraphics[scale=0.7]{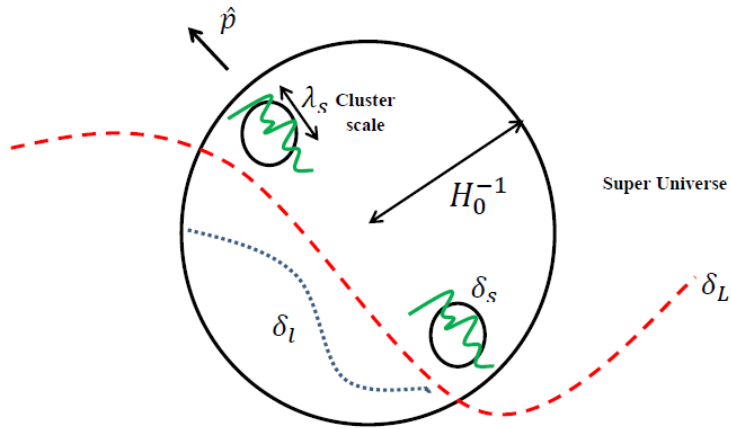}
\caption{This is a schematic figure representing  perturbation scales in peak-background splitting scenario. The large black-solid circle represents our observable Universe with radius  $H_0^{-1}$ which is situated inside the super Universe. The small black solid circles represent
the structures (i.e. galaxy clusters) in scale of Mpc. The green solid wavy curves represent the density contrasts in structure scale $\delta_s$. The blue dotted curve is the long mode perturbation $\delta_l$ inside the horizon. The red long dashed curve is the super-horizon mode $\delta_L$ in asymmetry direction $\hat{p}$.   }
\label{schem-fig}
\end{figure}

\begin{figure}
\includegraphics[scale=1.2]{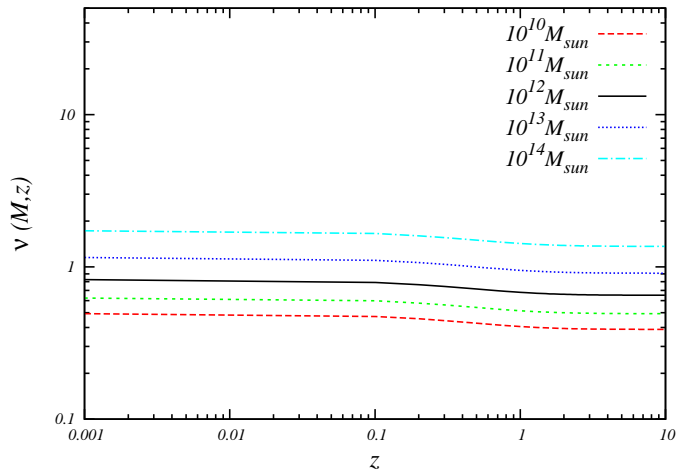}
\caption{In this figure we plot the height parameter $\nu=\delta_c/\sigma(M,z)$ versus redshift for different mass scales which are probed via LSS. The black solid line indicates the height parameter for $M=10^{12}M_{\odot}$, which is a typical mass scale of Luminous Red Galaxies (LRG)  }
\label{nu-fig}
\end{figure}
In order to study the effects of the long mode modulation on the statistics of the structures
we use the peak-background splitting method \cite{Bardeen:1985tr}, where the total matter density contrast $\delta \equiv \delta \rho_m/\rho_m$, is separated into the long and short wavelengths
\ba
\delta=\delta_s+\delta_l+\delta_L \, ,
\ea
where $\delta_s$ is the density contrast of the structure associated with the short wavelength,
$\delta_l$ is the density contrast from  the long wavelength mode in observable Universe  and $\delta_L$ represents the density perturbation originated from ${\cal{R}}_L$. (The different scales presented in this splitting is shown in Fig. \ref{schem-fig} ). The criteria to have  a structure is that the local density perturbation $\delta_s$ of structure passes the critical density threshold satisfying the condition
\ba \label{eq-critical}
\delta_s >\delta^{eff}_c\equiv \delta_c-\delta_l-\delta_L \, ,
\ea
where we have assumed that the scale of structures is much smaller than Hubble radius, $\lambda_s\ll H_0^{-1}$. Consequently, we assume that the effect of super-horizon long mode modulation is similar to the effect of long mode inside horizon. Both long modes, change the critical density by the amount calculated in Eq. (\ref{eq-critical}). Another important point to emphasis here is that we want to compare the statistics of structures in two different spatial positions where the amplitude of long mode modulation is slightly different. This means that the mean density of matter is almost the same inside the Hubble radius, while the perturbations are slightly different in dipole direction. 

The long wavelength mode changes the spherical collapse threshold so we expect to have a change in the statistics of the structures in the Universe. We would like  to calculate the probability of having structures with mass $M\equiv 4/3\pi R^3\rho_m$ (or equivalently the regions of radius $R$) and accordingly to calculate the bias parameter. To go further,  we define the effective height parameter as
\ba
\nu_{eff}\equiv\frac{\delta^{eff}_c}{\sigma(M)}=\nu- \frac{\delta_l+\delta_L}{\sigma(M)}
\ea
in which $\nu\equiv\delta_c/\sigma(M)$ and $\sigma(M)$ is the mass variance. The $\nu$ parameter plays a crucial role in the statistics of collapsed object. In Fig. \ref{nu-fig} we plotted the height parameter versus the redshift. For high mass objects in the Universe, the variance is lower, consequently the height parameter is larger ( meaning that their statistics is smaller). The redshift dependence of the height parameter comes from the growth function of structures. As the Universe becomes dark energy dominated the growth of the structures decreases, consequently we will have a small mass density and larger height function. The shift in height parameters occurred in $z\sim 0.3$ as in Fig. \ref{nu-fig}.

Now we can define the bias parameter from the modulated number density of the structures. The halo density contrast  is defined as
\ba
\delta_h=\frac{n(M,\nu_{eff})-\bar{n}(M)}{\bar{n}(M)}
\ea
where $\bar{n}(M)$ is the background number density of structures with mass $M$, and $n(M,\nu_{eff})$ is the modulated number density. Now the bias parameter is defined as
\ba
b=-\frac{1}{\sigma(M)}\frac{\partial \ln \bar{n}(M)}{\partial \nu}=-\frac{1}{\sigma(M)}\frac{f'}{f} \, ,
\ea
where we assumed the universality condition for the mass function of structures which means that they are only a function of the height parameter, $\bar{n}(M)\propto f(\nu)$. The prime here and below indicates the derivative with respect to $\nu$.  Now, the bias parameter becomes  a function of $\delta_L$, in contrast to the standard case.

Considering the fact that both $\delta_l$ and $\delta_L$ should be smaller than the critical  density and also $\delta_l\gg\delta_L$ so the effect of long mode modulation is sub-leading,   we can expand the bias parameter in terms of $\delta_L$ as
\ba
b=b_0+\frac{\partial b_0}{\partial\delta_L}\delta_L =b_0-\frac{1}{\sigma(M)}b_0'\delta_L \, ,
\ea
where $b_0$ represents the background unmodulated bias. Now we can go further and write the bias parameter in the terms of  universality function $f$ and $b_0$ as
\ba
b(\bx)=b_0+b_0^2\left(\frac{f''f}{f'^2}-1\right)\delta_L
\ea

Correspondingly, the gradient of bias associated with the long wavelength mode is obtained to be
\ba \label{nablab1}
\frac{\nabla b(\bx)}{b} = b_0\left(\frac{f''f}{f'^2}-1\right)\nabla\delta_L \equiv
b_0{\cal{F}}(\nu)\nabla\delta_L
\ea
in which
\ba
{\cal{F}}(\nu)\equiv \frac{f''f}{f'^2}-1 \, .
\ea

To go further, we can relate the long mode modulation of density perturbation to modulation in Bardeen potential via Poisson equation
\ba
\label{deltaL}
\delta_L=\frac{2}{3}\frac{\nabla^2 \Phi}{(1+z)H_0^2\Omega_m^0}\simeq \frac{2}{5}\frac{D(z)\nabla^2 {\cal{R}}_{L\, pri}}{H_0^2\Omega_m^0} = \bar{M}(z)\bar{\nabla}^2{\cal{R}}_{L\, pri}
\ea
where $D(z)$ is the growth factor,  $\Omega_m^0$
is the current fraction of matter energy density,  $\bar{M}(z)\equiv 2D(z)/5\Omega_m^0$,
and $\bar{\nabla}^2\equiv \nabla^2/H_0^2$. To obtain the approximate equality in
Eq. (\ref{deltaL}) we assumed that for the very long wavelength the transfer function
is unity,  $T(k)\simeq 1$ (i.e.  the growth of the potential function is scale-independent). In this case we use the relation   $\Phi=\frac{9}{10}(1+z)D(z)\Phi_{pri}$  in which $\Phi_{pri}$  represents the primordial value of $\Phi$ at the start of radiation (end of reheating) which is related to the primordial value of $R_L$ via $\Phi_{pri}=\frac{2}{3}{\cal{R}}_{L\, pri}$.  Consequently,  the gradient of the bias is obtained to be
\ba
\label{nablab2}
\frac{\nabla b(x)}{b}\simeq b_0{\cal{F}}\bar{M}(z)\nabla \left(\bar{\nabla}^2 {\cal{R}}_{L\, pri}\right)
\ea
Interestingly we see that the gradient in bias is related to the gradient in $\nabla^2 \calR_{L\, pri}$.
This analysis also shows another manifestation of the long wave length mode's effect on cosmological parameters. It shows that if the long mode modulation is the source of  dipole  asymmetry on CMB power spectrum, it will also  induce dipole asymmetry on LSS bias parameter. The long mode modulation introduces the asymmetry in bias parameter because of its effect on perturbations. As we have indicated in the Introduction, the  long mode modulation does not affect the background because of the re-scaling but it shows up at the perturbation level.

Let us define the amplitude of bias dipole asymmetry $A_b$ via
\ba
b=b_0\left(1+A_b\, \hat{\bf x}.\hat{\bf p}\right) \, .
\ea
As a result,  the gradient of the bias parameter is translated into
\ba
\label{grad-b}
\frac{\nabla b}{b_0} = \frac{ A_b \hat {\bf p}}{x_{LSS}} \, ,
\ea
where $x_{LSS}$ is the comoving distance from observer to the structure where  the bias parameter is measured.

So far our discussions were generic and model independent. Now we assume that the long mode modulation has the simple sinusoidal form as given in Eq. (\ref{RL}).
Combining  Eqs. (\ref{nablab2}) and (\ref{grad-b}),
the anisotropy bias parameter can be written as
\ba
\label{Ab-initial}
A_b\simeq b_0 {\cal{F}}(\nu)x_{LSS} \bar{M}(z)\nabla\left(\bar{\nabla}^2 {\cal{R}}_{L\, pri}\right)= b_0 {\cal{F}}(\nu) \bar{M}(z) \left(\frac{x_{LSS}}{x_{CMB}}\right)\left( \frac{k_L}{H_0}\right)^2 \left[ k_L x_{CMB}{\cal{P}}^{1/2}_{{\cal{R}}_L}\right]
\ea

\begin{figure}
\includegraphics[scale=1.2]{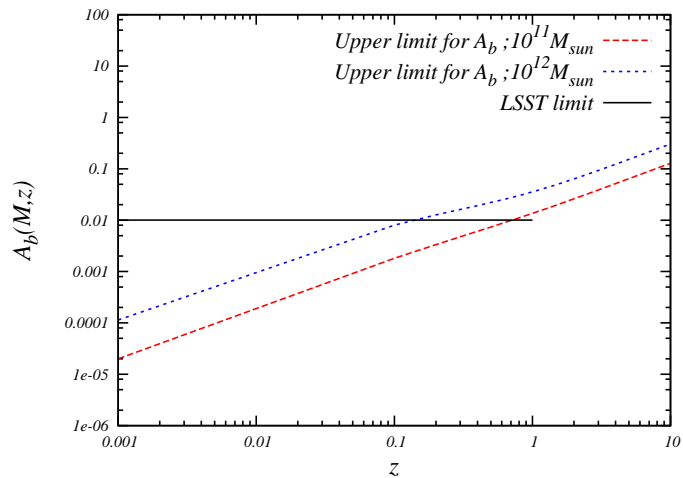}
\caption{In this figure we have plotted the bias anisotropy upper bound versus redshift, assuming the universal mass function of Press-Schechter formalism. We have set $M=10^{11}M_{\odot} (red long-dashed line)$ and $M=10^{13}M_{\odot}$ (blue dashed line)  and $k_L/H_0 = 1/2$. The  solid black horizontal line indicates the upper limit precision of LSST project. }
\label{Ab-fig}
\end{figure}
To simplify further,  we assume the initial conditions are Gaussian and the
probability function of structure formation has a universal form,  $n(M)\propto f(\nu)$. To be further specific, we  assume the
Press-Schechter universality function \cite{Press:1973iz} in which
{\ba
f(\nu)=\frac{\nu}{\sqrt{2\pi}} e^{-\nu^2/2}
\ea
and
\ba
{\cal{F}}(\nu)= -\frac{\nu^2+1}{(\nu^2-1)^2} \, .
\ea
For almost all observational cases where $\nu\ll 1$, ${\cal F}$ is a decreasing function of
$\nu$. Furthermore, ${\cal F}$ diverges at $\nu =1$, corresponding to  very high massive  cluster of galaxies (${\cal{O}}\sim 10^{14}M_{\odot}$). However,  because of the low statistics of the cluster of galaxies,  this is not suitable to obtain the bias parameter with the present and near future cosmological data.

Now  we can  use the observational constrain \cite{Namjoo:2013fka}  $k_L x_{CMB}{\cal{P}}^{1/2}_{{\cal{R}}_L }\leq 10^{-1}$  to put an upper bound on the bias parameter anisotropy. On the other hand the detection of dipole anisotropy in CMB $A\simeq 0.07$ and the assumption of non-Gaussianity at the  order of $f_{NL}\leq 10$ (which is compatible with Planck data) can be used to put an upper bound on anisotropic bias  as
\ba
\label{Ab-final}
 0.007 \times b_0 {\cal{F}}(\nu) \bar{M}(z) \left(\frac{x_{LSS}}{x_{CMB}}\right)\left( \frac{k_L}{H_0}\right)^2 \le A_b\le 10^{-1} \times b_0 {\cal{F}}(\nu) \bar{M}(z) \left(\frac{x_{LSS}}{x_{CMB}}\right)\left( \frac{k_L}{H_0}\right)^2 \, .
\ea
This is our main result in this section.

In a realistic  case,  taking  the galaxy samples  from SDSS - data release -9 \cite{Anderson:2012sa}, the mean redshift of the survey is set to $z=0.57$ which results in $\bar{M}\sim 1$ and $x_{LSS}/x_{CMB}\sim 2200 / 14000 \sim 0.15$.  Furthermore,   the linear bias parameter obtained from the Red Luminous Galaxies with $\nu\sim 0.8$ is $b_0\sim 0.8$ and $|{\cal{F}}|\sim 12.5$.  Taking the long mode to be, say, twice as big as the Hubble radius,  ${k_L}/{H_0}\sim 1/2$,
the upper bound on the amplitude of the bias dipole is obtained to be
\ba
A_b(z\simeq 0.5)\le 3\times  10^{-2} \, .
\ea
We have omitted our lower bound on asymmetric bias, because the long mode modulation could be much longer than the observable Universe and make the lower bound very small.
This  value is too small to be observed by today's large scale surveys but will be observable with the future large scale structure surveys like LSST. Indeed, this is within the error bar of the bias parameter and the errors originated from the peculiar velocity.  The Large Synoptic Survey Telescope (LSST)  which is designed to  obtain photometric redshift for 4 billion galaxies with the distribution peaking around z = 1,  can be used to determine the galaxy bias with high accuracy. The galaxy cluster count with combination of other cosmological observations, such as the  weak gravitational lensing and CMB data, can measure the bias parameter  in the redshift range between 0 to 1, with a precision as good as  2\% accuracy \cite{Zhan:2006gi}. Consequently, the anisotropy change in bias parameter must be greater than this error bar (systematic and statistic errors) to be detected in this redshift
range.

In Fig. \ref{Ab-fig} we plotted the bias anisotropy parameter versus redshift for two different mass scales, in which we calculated the bias parameter. The figure shows that in an optimistic case with high redshift $z\sim 1$, the upper bound of LSST is lower than the anisotropy calculated from a long mode with the size twice the size of the observable Universe.  In a optimistic situation, if one can measure the matter power spectrum in 21 cm in future observations such as SKA with $z\sim 10$, which results in $\bar{M}\sim 1.5$ and $x_{LSS}/x_{CMB}\sim 10000 / 14000 \sim 0.7$, and assuming  the same values for linear bias, ${\cal{F}(\nu)}$ and $k_L/H_0$, we obtain
\ba
A_b (z\simeq 10)\le 2 \times10^{-1} \, .
\ea
This is an ${\cal{O}}(20\%)$ change in the bias parameter. This is within the range of the observational detection.

\section{Dipole Asymmetry in Acceleration Expansion?}
\label{decel}

Having studied dipole asymmetries generated in tensor perturbations power spectrum and halo bias parameter, it is an interesting question if one searches for dipole asymmetries in late-time cosmological phenomena. In particular, we look into the possibility of generating dipole asymmetry in dark energy acceleration expansion induced from the long mode modulation. Somewhat related to this idea,  Kolb et al \cite{Kolb:2005me} have employed the idea of super-horizon perturbations to explain the late-time cosmological acceleration. But shortly after their proposal, it was shown that this idea does not work \cite{Geshnizjani:2005ce, Hirata:2005ei}. 

Employing the separate Universe approach, the line element in comoving gauge in the presence of long wavelength modulation is
\ba
ds^2 = -dt^2 + \bar a^{2}(t) e^{2 \calR_L(\mathbf{x},t)} \delta_{ij}dx^i dx^j \, ,
\ea
in which $a(t)$ is the background (average)  scale factor of our patch and $\calR$ denotes the curvature perturbation in comoving gauge. Again we do not mention the origin of this modulation or its shape but we just assume that this modulation can have slight variations in our observable patch. What we have in mind is that this modulation is the same which caused the hemispherical asymmetries in
CMB power spectrum, in tensor perturbations power spectrum and in bias
which were studied in previous Sections. In this view, all we need is that there exists a long mod modulation with large amplitude, i.e. ${\cal P}_{\calR_L} \gg {\cal P}_{\calR_{CMB}}$.

The above form of metric suggests that the effective scale factor in each Hubble patch is given by
\ba
\label{a-eff}
a(\bx, t) = \bar a(t) e^{\calR(\bx, t)} \, .
\ea
As a result, the effective Hubble expansion rate, $H= \dot a/a$,
is given by
\ba
\label{H-eff}
H(\mathbf{x},t) =  \bar H(t) + \dot{\calR}_L(\mathbf{x},t) \, ,
\ea
in which $\bar H(t)$ represents the background homogeneous Hubble parameter.
This equation suggests that the Hubble expansion rate is modified in the presence of the long wavelength mode.
It is important to note that if $\calR_L$ is time-independent then  it has  no effect on $H$ because in this limit $\calR_L$ can be absorbed into a re-scaling of $d \bx^i$ without affecting the expansion dynamics. Naively, one may imagine  that a mild variation of $\calR_L$ across the observable patch may result in a variation in $H$. Here we examine this idea critically for the case where the acceleration expansion is driven by a cosmological constant to see
whether or not $\calR_L$ can be time-dependent in late-time accelerating universe.

It is convenient to work with the deceleration parameter $q$ defined via
\ba
\label{q-def}
q \equiv - \frac{a \ddot a}{\dot{a}^2} \, .
\ea
With the effective scale factor and the Hubble parameter given in Eqs.  (\ref{a-eff}) and
(\ref{H-eff}), and to linear order in $\calR_L \ll 1$ and $ \dot{\calR}/H \ll 1$,  we get
\ba
\label{q-R}
q(\mathbf{x})\simeq \bar q -  \dfrac{2 \dot{\calR_L}(\mathbf{x})}{\bar H} (1+\bar q) - \dfrac{\ddot{\calR_L}(\mathbf{x})}{\bar H^2} \, ,
\ea
in which $\bar q$ represents the background homogeneous deceleration parameter in the absence of modulation.

To go further, we need to calculate $\calR_L(\bx, t)$ and its time derivatives in the late time accelerating expanding Universe and relate it to its primordial value at the end of inflation $\calR_{L\, pri}$. In order to find the time evolution of curvature perturbations, one can use the dynamical equation of the Bardeen potential $\Phi$. In general  the dynamical equation for the Bardeen potential can not be solved exactly. Usually solutions contain elliptic integrals even on the super horizon scales. Fortunately for the case at hand the situations simplify considerably. We consider the late-time universe containing only matter and cosmological constant. For both of these fluids, the perturbations in pressure is zero. On the other hand, the sound speed of fluctuations is defined in the comoving gauge via $\delta P_c = c_s^2 \delta \rho_c $. As a result, for both fluids the  sound speed of fluctuations is zero. Therefore, the effective sound speed of the total perturbations is zero too. In this limit one can show that there exists an invariant of dynamical equation. In addition, one can also show that this invariant
quantity coincides with the  gauge invariant curvature perturbation ${\cal R}$. To see this, consider  the evolution equation for the Bardeen potential   \cite{Mukhanov:book}
\ba
u''-c_s^2 \Delta u-\dfrac{\theta''}{\theta} u =0 \, ,
\ea
where
\ba
\theta \equiv \dfrac{1}{a} \left(1+ \dfrac{\bar{p}}{\bar{\rho}} \right) ^{-1/2} \, ,
\ea
and $u$ is defined as $u \equiv \Phi/ (\bar{\rho}+\bar{p})^{1/2}$.

As argued above, for our late-time universe containing only matter and cosmological constant $c_s=0$. In this limit one can easily manipulate the above equation to show that the quantity
$X$ defined via
\ba
X \equiv  \theta^2 \left(\dfrac{u}{\theta} \right)' \, ,
\ea
is a constant of integration. On the other hand, using the definitions of $u$ and $\theta$, and
noting that  ${\cal R} = \Phi - H/\dot{H} (\dot{\Phi}+H\Phi)$, one can easily show that
$\calR \propto X$. As a result,   $\calR$ is constant too. This indicate that there is no change in the deceleration parameter as given in Eq. (\ref{q-R}). This indicates that the long mode modulation does not induce asymmetry in acceleration expansion associated with the cosmological constant.

Having this said, it is possible to generate dipole asymmetry in late-time acceleration expansion if one considers other sources of dark energy. Our conclusion above was valid for cosmological constant in which the sound speed is zero. If one considers dynamical model of dark energy, such as the quintessence model, then the sound speed associated with the scalar field fluctuations are not zero. As a result,  in the model containing the mixed fluids of
matter and quintessence, $\calR$  can be time-dependent on super-horizon scales.
 It is an interesting question to see how the long mode modulation can generate dipole asymmetry in acceleration expansion for this scenario. However, this is beyond the scope of this work.

 Observationally, the possibility of anisotropic acceleration expansion is an intriguing idea though there are some controversies about the amplitude and the direction of dipole asymmetry in deceleration parameter, see e.g. \cite{Zhao:2013yaa} and  \cite{Cai:2013lja}. Some older works studied  this effect observationally \cite{Schwarz:2007wf}. Also recently, this issue has been revisited by many authors \cite{Zhao:2013yaa, Appleby:2012as, Cai:2013lja, Kalus:2012zu, Campanelli:2010zx}. Bonvin et al  showed that the dipole associated with  the luminosity distance is a useful observational tool which can be used to determine the Hubble parameter as a function of redshift H(z). They showed that our  peculiar velocity relative to CMB can induce dipole asymmetry on luminosity distance parameter. There are also some other works dealing with the effects of  peculiar velocities associated with the observers and supernovas
\cite{Davis:2010jq, Colin:2010ds, Hui:2005nm}. The typical correction to the  luminosity distance parameter due to the peculiar velocities can be estimated  to be $\Delta d_L /d_L \simeq v^{pec}/c$ \cite{Davis:2010jq, Hui:2005nm}. The typical peculiar velocities of
supernovas are at the order of $\sim 100 km/s$ which lead to $\Delta d_L /d_L \sim 10^{-3}$. One may expect that the dipole correction due to peculiar velocity of Earth to be in the same direction as the dipole of the CMB \cite{Bonvin:2006en} . But the observed dipole of the luminosity distance is at least one order of magnitude lager than expected asymmetry emerging from peculiar velocities \cite{Cai:2013lja}.  Moreover, there are also controversies on whether or not its dipole is  aligned with CMB dipole \cite{Zhao:2013yaa}. Recently, Zhao et  al claimed that they found a significant anisotropy with dipole amplitude $A_1=0.466^{+0.255}_{-0.205}$ which has  an angel $95.7^{\circ}$ with the
CMB dipole \cite{Zhao:2013yaa}.

\section{Conclusion and Discussions}
\label{conclusion}

The observed asymmetries in the CMB trigger the interest in super-horizon long wavelength modulations study as a probable explanation for the observed anomalies. In  \cite{Namjoo:2013fka} it was shown  that there is an upper bound consistency  relation between the anisotropy of CMB and the amount of local non-Gaussianity  induced by this long wavelength mode. The CMB temperature anisotropy moments  put strict constraints on the amplitude of anisotropy. In this work, following \cite{Namjoo:2013fka},
we presented a general formalism to investigate the consistency relation between the amplitude of dipole asymmetry and local non-Gaussianity induced by super-horizon mode. Then we have shown that this long mode power enhancement introduces a modulation on tensor perturbations. This suggests for further studies to check the possibility of tensor mode perturbation enhancement via CMB B-mode polarization which will be studied with more accuracy by Planck team next year.  We have obtained the consistency conditions Eqs. (\ref{A-O-R}) and (\ref{AT-AR2})  respectively in attractor and non-attractor single field inflationary models  for the ratio $A_T/A_\calR$. As discussed, the key point is that the tensor modes remain frozen after horizon crossing so any modulation of long mode on tensor perturbations are encoded at and near the time of horizon crossing. In this view, tensor perturbations are insensitive to features happing at the end or after inflation such as in curvaton scenario.

Having this said, non of these two single field  scenarios can produce detectable $A_T$. Single field slow-roll models fail to generate large enough  dipole asymmetry in CMB curvature perturbation power spectrum so their prediction  is $A_T \ll 0.03$. On the other hand, non-attractor model are able to  generate large enough dipole asymmetry in  CMB curvature perturbation power spectrum with large enough $f_{NL}^{loc}$ as given in Eq. (\ref{f-NLcs}). However, in non-attractor model it turns out that $\epsilon$  and $r$  are very small so from the consistency condition Eq. (\ref{AT-AR2}) we obtain $A_T$ is too small to be detectable.  Having this said, our
consistency condition Eq. (\ref{A-O-R}) should be viewed as a proof of concept, that in principle, a long mode modulation can yield dipole asymmetry in tensor perturbations
power spectrum. In practice, one has to look for multiple fields model of inflation to see if large observable $A_T$ can be generated.

Recently in \cite{Pajer:2013ana} the Maldacena's consistency condition that
$f_{NL} \sim 1-n_s$ has been revisited. It is argued that for the standard single field slow-roll models, considering all contributions, one finds $f_{NL}=0$ with corrections quadratic in $k_L$. Having this said, our results, as long as observational considerations are concerned, are intact. This is because, in order to have observable dipole asymmetry,  one has to go beyond the simple single field slow-roll inflation models \cite{Namjoo:2013fka}, such as non-attractor models, in which the argument of  \cite{Pajer:2013ana} does not apply.

As another interesting example, we have looked into effects of long mode modulation on halo bias parameter. Defining the bias dipole asymmetry  parameter, $A_b$, we found an upper bound on $A_b$ given by Eqs. (\ref{Ab-initial}) and (\ref{Ab-final}). We found that
$A_b \propto \nabla (\nabla^2 \calR_L)$. This implies that in general
the direction of bias dipole is different than the direction of CMB power spectrum dipole. Furthermore, the amplitude of bias dipole compared to the amplitude of CMB power spectrum is suppressed by the factor $f_{NL}\, ( x_{LSS}/x_{CMB})$. As a result, to get a higher value of $A_b$, one has to look at higher redshift structures  which have a higher value of
$ x_{LSSS}/x_{CMB}$. We argued that an ${\cal O}(20 \%)$ modulation in bias parameter can be obtained for matter power spectrum in 21 cm observations with $z \sim 10$. However, in general  the detection of anisotropy in bias parameter is more difficult  than the anisotropy in distances. This is because we need enough statistics (i.e. number of galaxies) in different directions in order to reduce the shot noise in galaxy power spectrum and investigate the change in bias parameter.

Finally, we have studied the effects of long mode modulation on the acceleration expansion.
We have shown that in $\Lambda CDM$ no dipole asymmetry in acceleration expansion is generated. This is because in a $\Lambda CDM$ background $\calR$ is conserved on super-horizon scales so the effects of $\calR$ can be absorbed into a constant coordinate transformation. To induce dipole asymmetry in acceleration one has to look for models in which dark energy is realized dynamically, i.e. in quintessence model, in which the sound speed of scalar perturbations is not zero and for the mixture of the matter and the dark energy
fluids   $\calR$ can evolve on super-horizon scales.


In early universe  cosmological backgrounds the long mode modulation can change the  Hubble expansion rate and introduce a dipole anisotropy on it. So it is natural to think that cosmological parameters which are related to the Hubble expansion are affected by the long mode modulations. For example, the observations which use the  BAO as a standard ruler will be affected because  the angular diameter distance and  the sound horizon of baryonic oscillations are modulated. The other interesting example is the study of Big-Bang Nucleosynthesis (BBN) in the presence of long mode modulation. As in standard BBN calculations  we have to compare the rate of nuclear  interactions with the rate of the Hubble expansion. The former is a property of particle physics and nuclear physics which are not affected by the long mode, while the Hubble expansion rate is affected by the long mode modulation. As a result, one expects to see dipole asymmetry in hydrogen and helium distributions. It will be an interesting question to study these predictions in details and compare them with the observations.

Our main goal in this work was to demonstrate that if the observed dipole asymmetry in CMB power spectrum is from the long mode modulation, then this effect will also show its fingerprints on different cosmological observables relevant to different cosmological histories. One can find the consistency conditions relating the amplitude of CMB dipole asymmetry to the dipole asymmetries induced in the power spectrum of tensor perturbations and  halo bias parameter. A detection or otherwise of any of these predicted asymmetries will have profound implications on inflation dynamics and also for the  possible pre-inflationary physics.

\acknowledgements
We would like to thank Razieh Emami, Hoda Ghodsi, Eiichiro Komatsu and Sadegh Movahed   for useful discussions. We are grateful to Amir Hossein Tajdini for assistances in calculating the  scalar-tensor-tensor cross correlation in the Appendix. We also thank the anonymous referee for the careful and insightful comments which improved the presentations and the results of the paper.

\appendix
\section{The analysis of $\langle \calR h_{ij} h_{kl}   \rangle $ for non-attractor model}
\label{App1}

In Section \ref{tensor} we have borrowed Maldacena's analysis \cite{Maldacena:2002vr}  for the three-point function  $\langle \calR h_{ij} h_{kl}   \rangle $ to obtain the relation between $A_T$ and $A_\calR$ in the attractor single field models as given in Eq. (\ref{A-O-R}).
 Here we generalize Maldacena's analysis to non-attractor inflation models in which
 $\calR$ is not conserved on super-horizon scales during the non-attractor phase.

 We consider the single field non-attractor models given by the following action \cite{Chen:2013eea}
 \ba
 S=  \int d^4 x \left[  \frac{M_P^2}{2} R + P(X, \phi) \right]  \, ,
 \ea
in which $X \equiv  -g^{\mu \nu} \partial_\mu \phi \partial_\nu \phi/2$. The above action is not the
most general action for the single field models, but it is generic enough
 which can shed light on the value of $\langle \calR h_{ij} h_{kl}   \rangle $ in the non-attractor scenarios. 

 To perform the perturbation analysis, we chose the comoving gauge in which $\delta \phi=0$
 and the tensor perturbations $h_{ij}$ are written as
 \ba
 h_{ij} = a^2(t) e^{2 \calR} \hat {h}_{ij} \, ,
 \ea
in which $\det  \hat {h}_{ij}=1$.  To second order, we have
\ba
\hat {h}_{ij} \simeq \delta_{ij} + \gamma_{ij} + \frac{1}{2} \gamma_{i l} \gamma_{l j} \, ,
\ea
in which  $\gamma_{ii} = \gamma_{ij, i}=0$, so the tensor perturbations are  transverse and traceless.

 We are interested in cubic action containing one scalar perturbation and two gravitons.
 We skip the details of this analysis and provide the final form of the cubic action
\ba
S_{\calR \gamma^2} = \int d^4 x \, \frac{\epsilon a^5}{2 c_s^2}  \dot \gamma_{ij}\dot \gamma_{ij}  \partial^{-2} \dot \calR \, ,
\ea
supplemented with the field redefinitions
\ba
\calR = \calR_c -\frac{1}{32}  \gamma_{i j} \gamma_{i j} + \frac{1}{16} \partial^{-2}
\left(  \gamma_{i j}   \partial^2 \gamma_{i j}  \right) \, ,
\ea
and
\ba
\label{gamma-new}
\gamma_{ij} = \gamma_{ij \, c} - \frac{a^2 \epsilon}{c_s^2} \left(\partial^{-2} \dot \calR_c  \right)  \dot \gamma_{ij}
\ea
Here $\epsilon = -\dot H/H^2$ is the slow-roll parameter and $c_s$ is the sound speed of
scalar perturbations.

Compared to attractor cases, there are some important modifications that one has to take into account. First, the wave-function of $\calR$ is different in non-attractor model
so one has to take this into account in the in-in integrals \cite{Chen:2013eea}. Second,  on super-horizon scales $\dot \calR$ is not negligible so, in contrast to  Maldacena's analysis \cite{Maldacena:2002vr},
one can not neglect the contribution of $\dot \calR$ in field redefinition Eq. (\ref{gamma-new}).

Putting all the contributions together, and skipping the details,
we obtain the following result for the correlation function between one scalar and two gravitons in the Fourier space
\ba
\label{shape-full}
\langle \calR_{\bk_1} \gamma_{\bk_2}^{s_2}  \gamma_{\bk_3}^{s_3}  \rangle
= &&(2 \pi)^3 \delta^3 \left(\sum \bk_i  \right)   \frac{P_{\calR \, k_1}(t_e)}{k_2^3 k_3^3} \epsilon (t_e) H^2  \delta_{s_2 s_3}  \left[ -\frac{c_s k_1^3}{2} + \frac{c_s}{2} k_1 (k_2^2 + k_3^2) + \frac{24 c_s k_1}{K^3} k_2^2 k_3^2\left( k_2 + k_3 - \frac{k_2 k_3}{K}  \right) \right.
\nonumber\\
  && \left. - \frac{8 k_2^2 k_3^2}{K}    - \frac{3 k_2^2 k_3^2}{c_s k_1 K} (k_2 - k_3)^2  \left(\frac{1}{c_s k_1} + \frac{1}{K} \right) + \frac{9}{K^2} k_2^2 k_3^2 (k_2 + k_3)  -\frac{12 c_s}{K^2} k_1 k_2^2 k_3^2
\right] \, .
\ea
Here $K \equiv c_s k_1 + k_2 + k_3$,  $s_2$ and $s_3$ are the graviton polarizations
and $t_e$ represents the time of end of non-attractor phase.   To simplify the analysis, we have assumed that $c_s$ and $H$ are nearly constant during inflation.

The above formula has the interesting property that the momentum associated with the scalar
perturbations, $k_1$, alway comes with a factor of $c_s$. Physically this make sense because for the scalar perturbations the moment of `` sound horizon '' is determined by $c_s k_1 = a H$ while for the tensor perturbations with momenta $k_2$ and $k_3$ it is the usual horizon crossing, $k_{2, 3} = a H$, which matters.

In the squeezed limit in which $k_1 \ll k_1\simeq k_3$, we get
\ba
\label{shape-squeezed}
\langle \calR_{\bk_1} \gamma_{\bk_2}^{s_2}  \gamma_{\bk_3}^{s_3}  \rangle
= (2 \pi)^3 \delta^3 \left(\sum \bk_i  \right)
\frac{P_{\calR \, k_1} (t_e)}{2 k_2^3 } \epsilon(t_e) H^2  \delta_{s_2 s_3}
\ea
Curiously, there were two leading terms proportional to $1/c_s^2$ (in the limit of small $c_s$)
which canceled each other  so Eq. (\ref{shape-squeezed})  represents the effects of the remaining sub-leading term which is independent of $1/c_s$ factor.

Comparing with the definition of $f_{NL}^{\calR h}$, given in Eq. (\ref{fNL}), and summing over the two polarizations $s_1, s_2$, we obtain
\ba
\frac{12}{5}f_{NL}^{\calR h}  =  2 \epsilon (t_e)
\ea
Noting that $2\epsilon(t_e) = - n_T$, our result for $f_{NL}^{\calR h}$ in non-attractor models  coincides with the result  obtained in Maldacena's analysis for attractor models, given by Eq. (\ref{Maldacena}).



  J.~F.~Donoghue, K.~Dutta and A.~Ross,
  ``Non-isotropy in the CMB power spectrum in single field inflation,''
  Phys.\ Rev.\ D {\bf 80}, 023526 (2009)
  [astro-ph/0703455 [ASTRO-PH]].

\bibitem{Wang:2013lda}
  L.~Wang and A.~Mazumdar,
  ``Small non-Gaussianity and dipole asymmetry in the CMB,''
  arXiv:1304.6399 [astro-ph.CO].

\bibitem{Erickcek:2009at}
  A.~L.~Erickcek, C.~M.~Hirata and M.~Kamionkowski,
  ``A Scale-Dependent Power Asymmetry from Isocurvature Perturbations,''
  Phys.\ Rev.\ D {\bf 80}, 083507 (2009)
  [arXiv:0907.0705 [astro-ph.CO]].

\bibitem{McDonald:2013aca}
  J.~McDonald,
  ``Isocurvature and Curvaton Perturbations with Red Power Spectrum and Large Hemispherical Asymmetry,''
  arXiv:1305.0525 [hep-ph].

\bibitem{Mazumdar:2013yta}
  A.~Mazumdar and L.~Wang,
  ``CMB dipole asymmetry from a fast roll phase,''
  arXiv:1306.5736 [astro-ph.CO].

\bibitem{Liu:2013kea}
  Z.~-G.~Liu, Z.~-K.~Guo and Y.~-S.~Piao,
  ``Obtaining the CMB anomalies with a bounce from the contracting phase to inflation,''
  arXiv:1304.6527 [astro-ph.CO].

\bibitem{Schmidt:2012ky}
  F.~Schmidt and L.~Hui,
  ``Cosmic Microwave Background Power Asymmetry from Non-Gaussian Modulation,''
  Phys.\ Rev.\ Lett.\  {\bf 110}, 011301 (2013)
  [Publisher-note {\bf 110}, 059902 (2013)]
  [arXiv:1210.2965 [astro-ph.CO]].

\bibitem{Prunet:2004zy}
  S.~Prunet, J.~-P.~Uzan, F.~Bernardeau and T.~Brunier,
  ``Constraints on mode couplings and modulation of the CMB with WMAP data,''
  Phys.\ Rev.\ D {\bf 71}, 083508 (2005)
  [astro-ph/0406364].

\bibitem{Byrnes:2011ri}
  C.~T.~Byrnes, S.~Nurmi, G.~Tasinato and D.~Wands,
  ``Inhomogeneous non-Gaussianity,''
  JCAP {\bf 1203}, 012 (2012)
  [arXiv:1111.2721 [astro-ph.CO]].

\bibitem{Kanno:2013ohv}
  S.~Kanno, M.~Sasaki and T.~Tanaka,
  ``A viable explanation of the CMB dipolar statistical anisotropy,''
  arXiv:1309.1350 [astro-ph.CO].

\bibitem{Maldacena:2002vr}
  J.~M.~Maldacena,
  ``Non-Gaussian features of primordial fluctuations in single field inflationary models,''
  JHEP {\bf 0305}, 013 (2003)
  [astro-ph/0210603].

\bibitem{Namjoo:2012aa}
  M.~H.~Namjoo, H.~Firouzjahi and M.~Sasaki,
  ``Violation of non-Gaussianity consistency relation in a single field inflationary model,''
  arXiv:1210.3692 [astro-ph.CO].

\bibitem{Chen:2013aj}
  X.~Chen, H.~Firouzjahi, M.~H.~Namjoo and M.~Sasaki,
  ``A Single Field Inflation Model with Large Local Non-Gaussianity,''
  arXiv:1301.5699 [hep-th].

\bibitem{Huang:2013oya}
  Q.~-G.~Huang and Y.~Wang,
  ``Large Local Non-Gaussianity from General Single-field Inflation,''
  arXiv:1303.4526 [hep-th].

\bibitem{Chen:2013kta}
  X.~Chen, H.~Firouzjahi, M.~H.~Namjoo and M.~Sasaki,
  ``Fluid Inflation,''
  arXiv:1306.2901 [hep-th].

\bibitem{Ade:2013ydc}
  P.~A.~R.~Ade {\it et al.}  [Planck Collaboration],
  ``Planck 2013 Results. XXIV. Constraints on primordial non-Gaussianity,''
  arXiv:1303.5084 [astro-ph.CO].

\bibitem{Hirata:2009ar}
  C.~M.~Hirata,
  JCAP {\bf 0909}, 011 (2009)
  [arXiv:0907.0703 [astro-ph.CO]].

\bibitem{Percival:2006gt}
  W.~J.~Percival, R.~C.~Nichol, D.~J.~Eisenstein, J.~A.~Frieman, M.~Fukugita, J.~Loveday, A.~C.~Pope and D.~P.~Schneider {\it et al.},
  ``The shape of the SDSS DR5 galaxy power spectrum,''
  Astrophys.\ J.\  {\bf 657}, 645 (2007)
  [astro-ph/0608636].

\bibitem{Mo:1995cs}
  H.~J.~Mo and S.~D.~M.~White,
  ``An Analytic model for the spatial clustering of dark matter halos,''
  Mon.\ Not.\ Roy.\ Astron.\ Soc.\  {\bf 282}, 347 (1996)
  [astro-ph/9512127].

\bibitem{Gunn:1972sv}
  J.~E.~Gunn and J.~R.~Gott, III,
  Astrophys.\ J.\  {\bf 176}, 1 (1972).

\bibitem{Baghram:2013lxa}
  S.~Baghram, M.~H.~Namjoo and H.~Firouzjahi,
  ``Large Scale Anisotropic Bias from Primordial non-Gaussianity,''
  arXiv:1303.4368 [astro-ph.CO].

\bibitem{Bardeen:1985tr}
  J.~M.~Bardeen, J.~R.~Bond, N.~Kaiser and A.~S.~Szalay,
  ``The Statistics of Peaks of Gaussian Random Fields,''
  Astrophys.\ J.\  {\bf 304}, 15 (1986).

\bibitem{Press:1973iz}
  W.~H.~Press and P.~Schechter,
  ``Formation of galaxies and clusters of galaxies by selfsimilar gravitational condensation,''
  Astrophys.\ J.\  {\bf 187}, 425 (1974).

\bibitem{Anderson:2012sa}
  L.~Anderson, E.~Aubourg, S.~Bailey, D.~Bizyaev, M.~Blanton, A.~S.~Bolton, J.~Brinkmann and J.~R.~Brownstein {\it et al.},
  ``The clustering of galaxies in the SDSS-III Baryon Oscillation Spectroscopic Survey: Baryon Acoustic Oscillations in the Data Release 9 Spectroscopic Galaxy Sample,''
  Mon.\ Not.\ Roy.\ Astron.\ Soc.\  {\bf 427}, no. 4, 3435 (2013)
  [arXiv:1203.6594 [astro-ph.CO]].

\bibitem{Zhan:2006gi}
  H.~Zhan,
  ``Cosmic tomographies: baryon acoustic oscillations and weak lensing,''
  JCAP {\bf 0608}, 008 (2006)
  [astro-ph/0605696].

\bibitem{Kolb:2005me}
  E.~W.~Kolb, S.~Matarrese, A.~Notari and A.~Riotto,
  ``Primordial inflation explains why the universe is accelerating today,''
  hep-th/0503117.

\bibitem{Geshnizjani:2005ce}
  G.~Geshnizjani, D.~J.~H.~Chung and N.~Afshordi,
  ``Do large-scale inhomogeneities explain away dark energy?,''
  Phys.\ Rev.\ D {\bf 72}, 023517 (2005)
  [astro-ph/0503553].

\bibitem{Hirata:2005ei}
  C.~M.~Hirata and U.~Seljak,
  ``Can superhorizon cosmological perturbations explain the acceleration of the Universe?,''
  Phys.\ Rev.\ D {\bf 72}, 083501 (2005)
  [astro-ph/0503582].

\bibitem{Zhao:2013yaa}
  W.~Zhao, P.~X.~Wu and Y.~Zhang,
  ``Anisotropy of Cosmic Acceleration,''
  arXiv:1305.2701 [astro-ph.CO].

\bibitem{Cai:2013lja}
  R.~-G.~Cai, Y.~-Z.~Ma, B.~Tang and Z.~-L.~Tuo,
  ``Constraining the Anisotropic Expansion of Universe,''
  arXiv:1303.0961 [astro-ph.CO].

\bibitem{Schwarz:2007wf}
  D.~J.~Schwarz and B.~Weinhorst,
  ``(An)isotropy of the Hubble diagram: Comparing hemispheres,''
  Astron.\ Astrophys.\  {\bf 474}, 717 (2007)
  [arXiv:0706.0165 [astro-ph]].

\bibitem{Appleby:2012as}
  S.~A.~Appleby and E.~V.~Linder,
  ``Probing Dark Energy Anisotropy,''
  Phys.\ Rev.\ D {\bf 87}, 023532 (2013)
  [arXiv:1210.8221 [astro-ph.CO]].

\bibitem{Kalus:2012zu}
  B.~Kalus, D.~J.~Schwarz, M.~Seikel and A.~Wiegand,
  ``Constraints on anisotropic cosmic expansion from supernovae,''
  Astron.\ Astrophys.\  {\bf 553}, A56 (2013)
  [arXiv:1212.3691 [astro-ph.CO]].

\bibitem{Campanelli:2010zx}
  L.~Campanelli, P.~Cea, G.~L.~Fogli and A.~Marrone,
  ``Testing the Isotropy of the Universe with Type Ia Supernovae,''
  Phys.\ Rev.\ D {\bf 83}, 103503 (2011)
  [arXiv:1012.5596 [astro-ph.CO]].

\bibitem{Cai:2011xs}
  R.~-G.~Cai and Z.~-L.~Tuo,
  ``Direction Dependence of the Deceleration Parameter,''
  JCAP {\bf 1202}, 004 (2012)
  [arXiv:1109.0941 [astro-ph.CO]].

\bibitem{Bonvin:2006en}
  C.~Bonvin, R.~Durrer and M.~Kunz,
  ``The dipole of the luminosity distance: a direct measure of h(z),''
  Phys.\ Rev.\ Lett.\  {\bf 96}, 191302 (2006)
  [astro-ph/0603240].

\bibitem{Davis:2010jq}
  T.~M.~Davis, L.~Hui, J.~A.~Frieman, T.~Haugbolle, R.~Kessler, B.~Sinclair, J.~Sollerman and B.~Bassett {\it et al.},
  ``The effect of peculiar velocities on supernova cosmology,''
  Astrophys.\ J.\  {\bf 741}, 67 (2011)
  [arXiv:1012.2912 [astro-ph.CO]].

\bibitem{Hui:2005nm}
  L.~Hui and P.~B.~Greene,
  ``Correlated Fluctuations in Luminosity Distance and the (Surprising) Importance of Peculiar Motion in Supernova Surveys,''
  Phys.\ Rev.\ D {\bf 73}, 123526 (2006)
  [astro-ph/0512159].

\bibitem{Colin:2010ds}
  J.~Colin, R.~Mohayaee, S.~Sarkar and A.~Shafieloo,
  Mon.\ Not.\ Roy.\ Astron.\ Soc.\  {\bf 414}, 264 (2011)
  [arXiv:1011.6292 [astro-ph.CO]].

\bibitem{Pajer:2013ana}
  E.~Pajer, F.~Schmidt and M.~Zaldarriaga,
  ``The Observed Squeezed Limit of Cosmological Three-Point Functions,''
  arXiv:1305.0824 [astro-ph.CO].

\bibitem{Mukhanov:book}
 V.~Mukhanov,
  ``Physical Foundations of Cosmology,''
  {\it Cambridge University Press} (2005).

\bibitem{Carroll:1991mt}
  S.~M.~Carroll, W.~H.~Press and E.~L.~Turner,
  Ann.\ Rev.\ Astron.\ Astrophys.\  {\bf 30}, 499 (1992).

 \bibitem{dodelson:2003}
 S.~Dodelson, \textit{Modern Cosmology} (Academic Press, 2003)

\bibitem{Chen:2013eea}
  X.~Chen, H.~Firouzjahi, E.~Komatsu, M.~H.~Namjoo and M.~Sasaki,
  ``In-in and $\delta N$ calculations of the bispectrum from non-attractor single-field inflation,''
  arXiv:1308.5341 [astro-ph.CO]; \\

\end{document}